\UseRawInputEncoding
\documentclass[letterpaper]{article} 
\usepackage{aaai2026}  
\usepackage{times}  
\usepackage{helvet}  
\usepackage{courier}  
\usepackage[hyphens]{url}  
\usepackage{graphicx} 
\usepackage{natbib}  
\usepackage{caption} 
\usepackage{algorithm}
\usepackage{algorithmic}
\usepackage{booktabs}
\usepackage{makecell}
\usepackage{subfig}
\usepackage{tikz}
\usepackage{tcolorbox}
\usepackage{amsfonts}
\usepackage{amsmath} 
\usepackage{fontawesome}
\usepackage[toc,page]{appendix}
\usepackage{newfloat}
\usepackage{listings}
\DeclareCaptionStyle{ruled}{labelfont=normalfont,labelsep=colon,strut=off} 
\floatstyle{ruled}
\newfloat{listing}{tb}{lst}{}
\floatname{listing}{Listing}
\title{From Time and Place to Preference: \\LLM-Driven Geo-Temporal\thanks{Geo-temporal combines temporal and geographic signals.} Context in Recommendations}

\author{
    Yejin Kim\textsuperscript{\rm 1,2},
    Shaghayegh Agah\textsuperscript{\rm 1},
    Neeraj Sharma\textsuperscript{\rm 1},
    Mayur Nankani\textsuperscript{\rm 1},\\
    Feifei Peng\textsuperscript{\rm 1},
    Maria Peifer\textsuperscript{\rm 1},
    Sardar Hamidian\textsuperscript{\rm 1}\footnotemark[2]
    H. Howie Huang\textsuperscript{\rm 2}\footnotemark[2]
}
\affiliations{
    \textsuperscript{\rm 1}Comcast Technology AI \\
    \textsuperscript{\rm 2}GraphLab, George Washington University \\
    \{yejin\_kim, shaghayegh\_agah, neeraj\_sharma, mayur\_nankani, feifei\_peng, maria\_peifer, sardar\_hamidian\}@comcast.com \\
    \{yeijnjenny, howie\}@gwu.edu \\
}

\usepackage{bibentry}

\begin{document}

\maketitle

\makeatletter
\renewcommand{\@makefnmark}{}
\makeatother
\footnotetext{
  \textdagger\ Corresponding authors.}
  
\begin{abstract}
Most recommender systems treat timestamps as numeric or cyclical values, overlooking real-world context such as holidays, events, and seasonal patterns. We propose a scalable framework that uses large language models (LLMs) to generate geo-temporal embeddings from only a timestamp and coarse location, capturing holidays, seasonal trends, and local/global events. We then introduce a geo-temporal embedding informativeness test as a lightweight diagnostic, demonstrating on MovieLens, LastFM, and a production dataset that these embeddings provide predictive signal consistent with the outcomes of full model integrations. Geo-temporal embeddings are incorporated into sequential models through (1) direct feature fusion with metadata embeddings or (2) an auxiliary loss that enforces semantic and geo-temporal alignment. Our findings highlight the need for adaptive or hybrid recommendation strategies, and we release a context-enriched MovieLens dataset to support future research.

\end{abstract}


\section{Introduction}

While many recommender systems incorporate temporal signals to capture periodic and seasonal patterns, they often fail to encompass the full spectrum of geo-temporal factors that shape user preferences. In practice, user intent is influenced not only by time but also by cultural, seasonal, and regional variations. For example, a user's preferences during winter in New York may differ substantially from those in Florida, particularly for multinational platforms that must serve a geographically and culturally diverse user base. Traditional models attempt to account for such variations by leveraging extensive user histories and manually engineered temporal features; however, these approaches face significant scalability challenges in real-world large-scale recommendation systems ~\cite{bogina2022temporal}.
\begin{figure}[t]
\centering
	\includegraphics[width=0.9\linewidth]{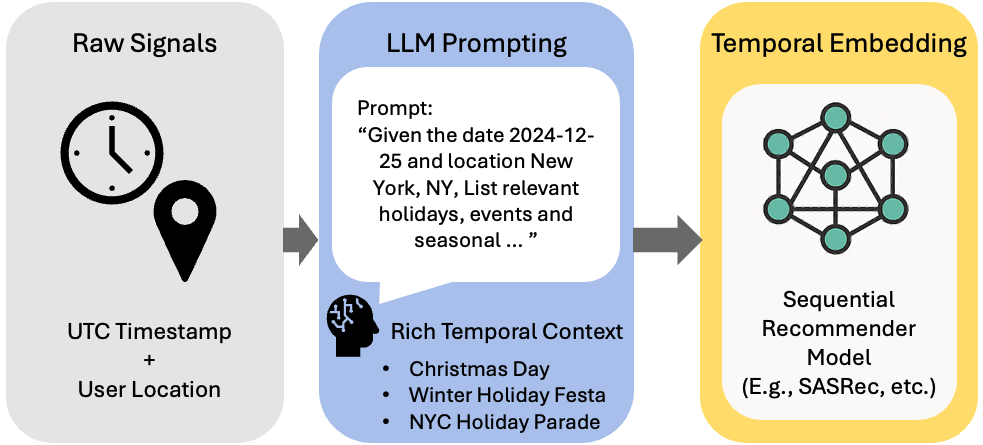}
\caption{From timestamp and location to LLM-generated geo-temporal context for sequential recommendation.
}
\label{fig:intro}
\end{figure}

Moreover, the impact of contextual information varies across domains. In e-commerce, fashion cycles and seasonal trends are dominant~\cite{ludewig2018fashion}, while in streaming platforms, entertainment events such as the Oscars or the Grammys can reshape consumption patterns~\cite{adomavicius2011context}. However, existing context-aware recommenders typically rely on static, domain-specific calendars or curated event lists~\cite{mateos2024systematic}, which limits responsiveness to real-time changes. Dynamic context modeling has emerged as a solution, but current methods demand substantial engineering effort, rely on curated external data, and often lack cross-domain generality~\cite{chen2023tourpie}.



Large language models (LLMs) now offer the capability to generate semantically rich, real-time contextual signals directly from coarse inputs such as timestamps and geographic locations. This opens the door to scalable, domain-agnostic personalization without manual feature engineering. Motivated by this, we propose an LLM-driven geo-temporal (GT) context enrichment framework that synthesizes cultural, seasonal, and event-based information into geo-temporal embeddings for recommendation. Figure~\ref{fig:intro} illustrates the process, from absolute timestamps and coarse location inputs to LLM-generated geo-temporal context features that are integrated into sequential recommendation models.

While LLM-generated context raises challenges such as hallucinations, outdated knowledge, and limited traceability, we demonstrate that these embeddings can deliver measurable performance gains in realistic recommendation scenarios. Our contributions are summarized as follows:
\begin{itemize}
    \item We propose a novel LLM-driven framework for dynamic temporal and location context enrichment in recommendation systems, extracting rich semantic signals from only a UTC timestamp and a coarse location without relying on static calendars or manual feature engineering.
    \item We introduce a geo-temporal embedding informativeness test as a lightweight diagnostic, and validate across MovieLens~\cite{harper2015movielens}, LastFM~\cite{Bertin-Mahieux2011}, and the production dataset that aligns with the corresponding evaluation results.
    \item We evaluate our proposed frameworks on two datasets, MovieLens 1M and a large-scale production dataset by integrating the generated geo-temporal context into multiple sequential recommendation variants, showing dataset- and behavior-dependent gains.
    \item We release context-enriched versions of MovieLens 1M with detailed geotemporal annotations to support research on adaptive, time- and event-aware personalization.
\end{itemize}

\section{Related Work}
Context-aware recommender systems integrate additional contextual information such as time, location, or social setting to better personalize recommendations. Temporal context is among the most important dimensions, as user preferences fluctuate based on day, season, or ongoing events. Studies have shown that incorporating time-based context improves both trust and engagement \cite{panniello2016trust, gorgoglione2019temporal}. 

Two primary approaches to encode time: (1) explicit feature-based modeling, where temporal information is incorporated as static metadata (e.g., hour of day, holiday flag), and (2) implicit sequence modeling, which captures temporal dynamics through evolving user-item sequences \cite{chiroma2017cars, raza2019overview}. Surveys such as \cite{bogina2022temporal} further emphasize that understanding seasonality, temporal drift, and context evolution is critical to performance across domains. Domain-specific studies confirm its value in e-commerce benefits from modeling seasonal promotions \cite{ma2019seasonality}, food recommenders from daily and weekly cycles \cite{trattner2019food}, and news platforms from global events \cite{gulla2016news}. However, most rely on handcrafted or rigid rules, limiting scalability.


Recent works in ubiquitous and mobile computing have enriched recommendation pipelines with dynamic context such as weather, user mobility, and physical environment. Yoon and Choi \cite{yoon2023tourism} showed how tourism recommenders benefit from incorporating live environmental signals (e.g., temperature or precipitation). Despite advances, few works model high-level semantic temporal knowledge such as “Christmas,” “World Cup,” or “Back to School season” as first-class contextual entities, largely due to limited knowledge integration in traditional recommender systems.

\subsubsection{LLMs for Contextual Enrichment}

Efforts to bridge this gap increasingly rely on external knowledge. Early work integrated structured databases (e.g., Wikidata) via knowledge graphs, enriching item and user representations with semantic attributes \cite{wang2018ripplenet}, improving explainability and cold-start handling. Unlike curated knowledge bases, LLMs expand this by encoding vast world knowledge and generating dynamic context. Eghbalzadeh et al.~\cite{eghbalzadeh2024ads} use a multimodal LLM to detect and calibrate seasonal trends in advertisements. Xu et al.~\cite{xu2025intent} show that LLMs can be used to refine or reinterpret user intent in recommendation tasks. LLMs have also been explored for augmenting interaction data \cite{wei2024synthetic} or generating item descriptions for improved personalization.



Our work leverages this trend to extract semantically rich geo-temporal context from timestamps and coarse location data. Instead of hard-coded holiday lists or domain-specific heuristics, we prompt an LLM to identify real-world events, cultural phenomena, or behavioral trends that coincide with user activity. This enables dynamic, scalable, and domain-agnostic enrichment of recommender models with real-world awareness, aligning with the broader push toward foundation model-augmented personalization systems.

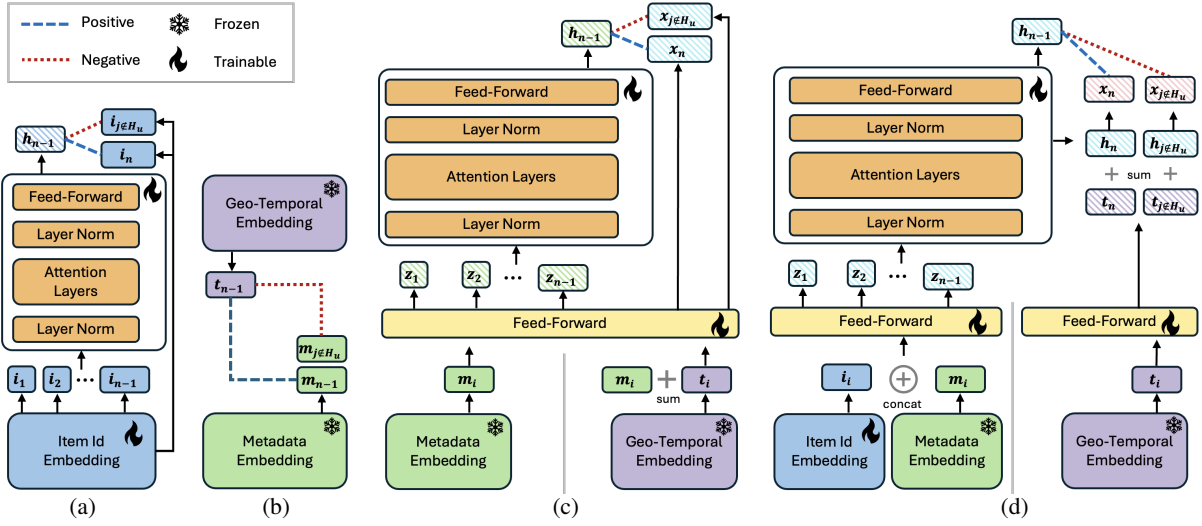
\begin{figure*}[t]
\centering
\begin{tikzpicture}
  \node[anchor=south west,inner sep=0] (image) at (0,0)
    {\includegraphics[width=0.9\linewidth]{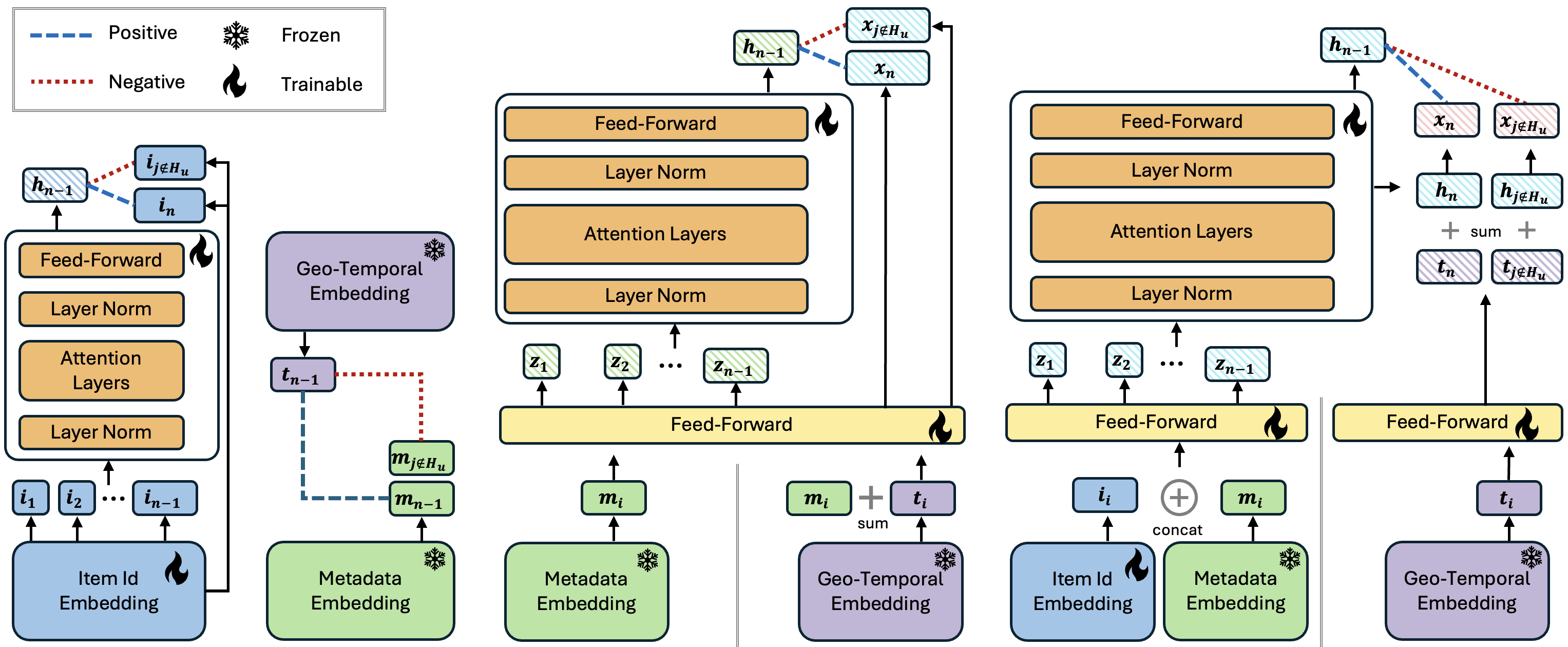}};
  \begin{scope}[x={(image.south east)},y={(image.north west)}]
    \node[anchor=north west] at (0.052,0.01) {\small\text{(a)}};
    \node[anchor=north west] at (0.212,0.01) {\small\text{(b)}};
    \node[anchor=north west] at (0.453,0.01) {\small\text{(c)}};
    \node[anchor=north west] at (0.825,0.01) {\small\text{(d)}};
  \end{scope}
\end{tikzpicture}
\caption{(a) Baseline model (SASRec) (b) Auxiliary loss contrasting semantically and geo-temporally similar \textit{vs.} dissimilar items (c) Replaced item Id embeddings with metadata embeddings as input, and added geo-temporal embeddings to metadata embeddings at the loss stage (d) Concatenated item Id and metadata embeddings as input, and added geo-temporal embeddings at the loss stage}
\label{fig:combined}
\end{figure*}

\section{Approach}
We propose a large language model (LLM) driven \textit{Geo-Temporal Context Extraction Module} that transforms a UTC timestamp and user location into a semantically rich set of contextual attributes. We define context as the set of geo-temporally relevant signals such as holidays, events, or cultural phenomena, that may influence user preferences at the time of interaction.

Given a timestamp and a geographic location (e.g., city, state, or country), we construct a structured prompt to query different LLMs ~\cite{meta2024llama3}. 
The prompt asks the model to describe the geo-temporal context relevant to that specific time and place, including seasonality, day type (weekday/weekend/holiday), ongoing cultural or regional events, global news, and behavioral trends. The LLM is instructed to return its response in a structured JSON format, which we parse into machine readable features suitable for downstream processing.

We incorporate these geo-temporal context features into a sequential recommendation model, alongside collaborative signals and item metadata features. We explore multiple integration strategies for fusing context into the ranking pipeline. Through these experiments, we aim to identify effective patterns for leveraging LLM-generated geo-temporal context in recommendation tasks.

\subsection{Prompt Design for Geo-Temporal Context}\label{prompt_design}

To extract rich geo-temporal context from LLMs, we design a structured prompt template grounded in the interaction timestamp and user location. The prompt instructs the LLM to return:

\begin{itemize}
    \item A list of relevant real-world events that occurred shortly before the specified timestamp in the given region, including holidays, cultural celebrations, political events, sports tournaments, entertainment releases, fashion trends, notable figures, and major social events.
    \item A concise natural language summary (2--3 sentences) describing the broader geo-temporal context and how it might influence user behavior.
\end{itemize}

The prompt enforces a structured JSON response with two fields: \small{\texttt{events}} (an array of event descriptions) and \small{\texttt{summary}} (a short explanatory text).

\begin{tcolorbox}[colback=gray!5!white,colframe=gray!75!black,title=Abstract Version of the Prompt]
\scriptsize
\texttt{You are a time-context assistant for recommendation systems.\\
A user interacted with content at the following time and place:\\
- UTC time: \{timestamp\}\\
- Location: \{location\}\\
Please return:\\
1. A list of relevant real-world events shortly before this time and location.\\
2. A short natural language summary explaining how these events might influence user interests.\\
Format your answer as valid JSON with fields \texttt{events} and \texttt{summary}.
}
\end{tcolorbox}
\normalsize



\subsection{Geo-Temporal Embedding Informativeness Test}
Before integrating geo-temporal embeddings into the full recommendation pipeline, we conduct an initial utility test to assess whether they encode meaningful predictive signals.

Let ${t}_i \in \mathbb{R}^d$ denote the geo-temporal embedding of item $i$ generated by the LLM from its interaction timestamp and location, and let ${h}_{u}$ represent the set of items previously interacted with by user $u$. For each user, we randomly sample one item $i \in {h}_{u}$ and use its temporal embedding ${t}_i$ as the \emph{user geo-temporal embedding}.

Let ${m}_j \in \mathbb{R}^d$ denote the metadata embedding of candidate item $j$, where $j$ ranges over the entire set. The user–item score is computed as the dot product:

\small
\begin{equation}
\mathrm{score}(u, j) = {t}_i^\top {m}_j
\end{equation}
\normalsize
Items are ranked in descending order of $\mathrm{score}(u, j)$ to produce recommendations. To evaluate the informativeness of geo-temporal embeddings, we measure the hit rate $\mathrm{HR}@k$ for this ranking and compare it against a random ranking baseline:
\small
\begin{equation}
\mathrm{HR}@k_{\mathrm{geo-temporal}} > \mathrm{HR}@k_{\mathrm{random}}
\end{equation}
\normalsize
If the above condition holds, even by a small margin, we conclude that the geo-temporal embeddings contain predictive information that can enhance recommendation quality. This serves as a lightweight validation step before integrating geo-temporal embeddings into more complex sequential models.






\subsection{Model Integration}

We integrated the LLM-generated geo-temporal context features into the sequential recommendation model SASRec~\cite{kang2018self}. To represent the context information, we encoded the JSON-based descriptions using the GTE-large encoder~\cite{li2023towards}, producing dense embeddings that were incorporated into SASRec’s input sequence alongside traditional item embeddings. Our goal was to evaluate not only the overall impact of geo-temporal context on recommendation performance, but also to explore effective strategies for integrating geo-temporal context features into existing architectures.

We experimented with multiple integration schemes, feature representations, and model configurations to identify practical design patterns for incorporating context-aware signals into sequential models. Figure~\ref{fig:combined} illustrates the architectures of our proposed context-enhanced sequential recommendation model, which extends a SASRec-style Transformer backbone by integrating LLM-generated geo-temporal context representations. 


In Figure~\ref{fig:combined}-(a), a SASRec encoder, serving as the baseline Transformer backbone, processes item Id embeddings through stacked self-attention and feed-forward layers to produce collaborative representations for next-item prediction. Each token corresponds to a learnable item Id embedding $i_{i}\in\mathbb{R}^{d_i}$, where the user’s history is $H_{u} = \{ i_{1}, i_{2}, \ldots, i_{n-1} \}$. Next-item relevance is scored by dot products with positive and negative items:

\small
\begin{equation}
s^{+} = h_{n-1}^\top i_{n}, 
\qquad 
s^{-} = h_{n-1}^\top i_{j}, \; j \notin H_{u}
\end{equation}
\normalsize
and training is guided by a contrastive loss:
\small
\begin{equation}\label{eq4}
\mathcal{L}_{rank} = -\log \sigma(s^{+}) - \log \sigma(-s^{-})
\end{equation}
\normalsize
We consider two complementary strategies for incorporating geo-temporal embeddings. One strategy introduces an auxiliary loss that encourages geo-temporal embeddings to align with semantic metadata representations (as in Figure \ref{fig:combined}-(b)). The other strategy integrates geo-temporal signals directly into the input sequence, either by summing them with metadata embeddings or combining them with Id embeddings (as in Figures \ref{fig:combined}-(c) and (d)).

Figure~\ref{fig:combined}-(b) shows an auxiliary loss aligning geo-temporal embeddings with item metadata embeddings. Given a geo-temporal embedding ${t}_{n-1}$ from the interaction timestamp and a metadata embedding ${m}_{n-1}$ of the same item, we treat them as a positive pair. In contrast, ${t}_{n-1}$ and ${m}_{j}$, where $j \notin H_{u1}$, form negative pairs. To construct negative samples, the method identifies items that the target user has not interacted with. For each interaction, it considers a geo-temporal window preceding the interaction date and gathers items interacted with by other users during that period. It then filters out any items the target user has already seen to avoid false negatives. From the remaining pool, a fixed number of items are randomly selected as negative samples—representing plausible alternatives the user could have encountered but did not engage with. The auxiliary objective encourages similarity for positives and dissimilarity for negatives:
\small
\begin{equation}
\mathcal{L}_{aux} = \ell\big(\mathrm{sim}({t}_{n-1}, {m}_{n-1}), 1\big)+
\ell\big(\mathrm{sim}({t}_{n-1}, {m}_{j}), 0\big),
\end{equation}
\normalsize
where $\mathrm{sim}(\cdot)$ denotes a similarity function and $\ell(\cdot)$ is a general contrastive loss. This auxiliary supervision encourages geo-temporal embeddings to capture semantic consistency with item content while remaining discriminative against unrelated items. Since both geo-temporal and metadata embeddings are precomputed and fixed, the auxiliary objective is defined purely at the embedding level. As a result, it is model-agnostic and can be seamlessly combined with a wide range of sequential encoders beyond SASRec.


Figure~\ref{fig:combined}-(c) depicts a variant that removes trainable Id embeddings and instead relies on metadata representations ${m}_i$, which replace the Id embeddings in the SASRec encoder of Figure~\ref{fig:combined}-(a). 
Temporal embeddings ${t}_i$ are combined with metadata embeddings during the scoring stage when constructing positive and negative pairs for the contrastive loss. Next-item relevance is computed as the dot product between the user representation, $h_{n-1}$, and positive/negative geo-temporal metadata embeddings:
\small
\begin{align}
h_{n-1} &= \mathcal{G}(\mathcal{F}([m_{1}, \ldots, m_{n-1}])), \quad x_{i} = \mathcal{F}(m_i + t_i), \\
s^{+} &= h_{n-1}^\top x_n, \qquad 
s^{-} = h_{n-1}^\top x_j, \quad j \notin H_{u} \label{eq7}
\end{align}

\normalsize
Equations~\eqref{eq7} are applied to Equation~\eqref {eq4} to compute the ranking loss. Since both ${m}_i$ and ${t}_i$ are fixed, the parameter count excludes the Id table and any additional projection layers, leaving only the Transformer backbone trainable. This makes (c) highly parameter- and compute-efficient, with negligible per-token overhead before the Transformer.

At inference, we simply calculate
$x_j = \mathcal{F}(m_j + t_j)$ (with context) or $x_j = \mathcal{F}(m_j)$ (without context), 
and the relevance score is $s_j = h_{n-1}^\top x_j$.

Figure~\ref{fig:combined}-(d) instead retains trainable Id embeddings and concatenates them with metadata embeddings, which are first projected into the model space; geo-temporal signals are then incorporated only at the ranking loss stage.
\small
\begin{align}
h_{n-1} = \mathcal{G}([z_{1},...,z_{n-1}]), \quad z_i = \mathcal{F}_1([\,i_i \,\|\, m_i\,])\label{eq8}\\
\quad x_{i} = \mathcal{F}_2(t_i) + h_{i} \qquad\qquad\quad \label{eq9}
\end{align}
\normalsize
Equations~\eqref{eq8} and \eqref{eq9} are applied to Equations~\eqref{eq7} and ~\eqref{eq4} to compute the ranking loss. In this design, a trainable Id embedding $i_i$, initialized from a lookup table, is concatenated with the frozen metadata embedding $m_i$ and projected into the model space. Geo-temporal embeddings $t_i$ (frozen from a pre-trained encoder) are then added only at the ranking loss stage. This preserves collaborative signals through $i_i$ while allowing metadata and geo-temporal features to act as contextual modulators. By deferring geo-temporal fusion, $t_i$ functions as a contextual bias rather than being entangled with Id and metadata at the input stage, which is beneficial when geo-temporal factors modulate but do not override co-consumption patterns uniquely captured by Id embeddings.

At inference, each candidate is scored as $x_j = h_j + \mathcal{F}_2(t_j)$ (with context) or $x_j = h_j$ (without context), with the relevance score $s_j = h_{n-1}^\top x_j$.

\section{Experiments}

\subsection{Dataset and Contextual Enrichment}
We selected the MovieLens 1M and LastFM 1K datasets due to their inclusion of both user interaction timestamps and location information, which are essential for generating location and time-specific context. It was necessary to have a dataset collected in a wide time window to capture seasonality in different contextualized scenarios. We also utilized a recent large-scale production dataset containing real user interaction logs to validate the practical utility of our approach in real-world systems. We applied a standard prompt-engineering pipeline to enrich these datasets, employing large language models (LLMs) to generate structured JSON outputs containing real-world events and natural language summaries of geo-temporal context. 

For evaluation, we constructed two types of test sets: a \textit{general user} setting, which reflects typical historical interactions and includes behaviors such as binge watching, and an \textit{explorer user} setting, where users interacted only with entirely new items not seen in the previous six months. This split allows us to examine both standard recommendation quality and the model’s ability to generalize to unseen items. In the production dataset, we had the option to retain or remove repeated series episodes. For the general user setup, we kept them based on total watched duration, while for the explorer user setup, we removed such duplicates during training.
\textit{Note that MovieLens timestamps reflect rating events, not actual watch times, adding noise from rating delays. Each movie appears only once per user, so all users are effectively explorer users.}
\vspace{-0.25cm}

\subsection{Evaluating Geo-Temporal Embedding}
To assess the predictive power of LLM-generated contextual features independently of user history, we conducted a non-parametric end-to-end (E2E) experiment. For randomly selected target days from the dataset, we embedded the contextual signal and computed cosine similarity with item representations derived from metadata, encoded using GTE-large~\cite{li2023towards}. Compared to a random baseline, contextual signals alone yielded meaningful performance, as measured by hit rate (HR), demonstrating their standalone predictive utility. 

\begin{table}[ht]
\centering
\caption{Percentage of improvements (\%) of HR@$k$ on three datasets as $(\text{HR}_{E2E} - \text{HR}_{Random}) / \text{HR}_{\text{Random}} \times 100$}
\label{e2e}
\resizebox{0.78\linewidth}{!}{%
\begin{tabular}{@{}lcccc@{}}
\toprule
\textbf{Dataset} & \textbf{@10} & \textbf{@20} & \textbf{@50} & \textbf{@100}\\ \midrule
MovieLens & +78.9 & +20.7 & +41.1 & +14.2 \\
Large-scale Production & +16.3 & +16.1 & +21.9 & +25.4 \\
LastFM & +0.79 & +0.39 & -0.30 & -0.61 \\
\bottomrule
\end{tabular}%
}
\end{table}
 
While MovieLens and the large-scale production dataset show clear performance gains, LastFM results are small and inconsistent in Table \ref{e2e}, indicating little geo-temporal signal for this domain. Consistently, full model integration on LastFM also yielded no meaningful improvements. We therefore report LastFM only as a preliminary validation of our diagnostic test, excluding it from the main results.



\vspace{-0.25cm}
\subsection{Evaluation Metrics}
We conducted a series of experiments to evaluate the performance of GT-enhanced recommendation model. Standard top-$k$ ranking metrics, hit rate (HR) and normalized discounted cumulative gain (NDCG), were used to assess the quality of the recommendation. For the \textit{general user} test set, we additionally report item coverage to capture the diversity of recommended items. In contrast, for the \textit{explorer user} setting, where users interact only with entirely new items, we focus on HR and NDCG, since the primary concern is ranking quality among unseen items rather than coverage.

\subsection{Model Variants and Terminology} We consider several model variants in our evaluation, M denotes the metadata-only baseline, and Id+M augments this with item Ids. Id+M+GT extends Id+M with geo-temporal features applied at both training and inference, while Id+M+GT$_{train}$ uses geo-temporal features only during training. Similarly, M+GT incorporates geo-temporal features into the metadata-only setting. In addition, we examine extensions with auxiliary objectives, denoted as +Loss-[BCE/Cos/Pairwise$_{\{rand|sem\}}$]. These variants correspond to the architectural designs in Figure~\ref{fig:combined}: (a) the metadata-only baseline, which yields M and Id+M depending on the input; (b) the auxiliary-loss extension (+Loss-*); (c) the metadata with geo-temporal features (M+GT); and (d) the full geo-temporal model (Id+M+GT). In Pairwise$_{sem}$, we use a customized pairwise loss based on Margin Ranking Loss and semantic negative samples. Negatives are sampled randomly from the pool of least similar items to the GT feature. We perform several ablation studies on the selection of negative samples for each GT embedding. 
Our studies demonstrated that restricting to the most dissimilar item from the entire space tends to have a lower performance compared with randomly selecting from the pool of dissimilar items. The latter yields more stable results. This can be due to overfitting of the pairwise loss on extremely hard negative samples.

\vspace{-0.25cm}
\subsection{Main Results}
For \textit{general users}, Table~\ref{tab:model_comparison production dataset random users} shows that M+GT achieves the strongest performance. By capturing next-item signals from both content information (genres, actors, topics) and geo-temporal effects (weekends, holidays, seasons), M+GT provides a lightweight and robust encoder. Since metadata and geo-temporal embeddings are fused while frozen, the model avoids reliance on collaborative memorization. This design also improves parameter efficiency, reducing model size and acting as a regularizer. As a result, M+GT yields better average accuracy across broad user groups by lowering variance and mitigating overfitting, making it particularly effective when content and geo-temporal signals dominate.  

For \textit{explorer users}, Table~\ref{tab:model_comparison_two_datasets explorer users} shows that 
Id+M+GT model categories deliver superior performance. These users exhibit high-entropy, novelty-seeking, or long-tail behavior where fine-grained, item-to-item cues (e.g., sequel chains) are essential. Unlike M+GT, Id+M+GT retains item IDs along with metadata and GT features. This provides stronger expressiveness and allows the model to capture fine-grained item-to-item transitions, leading to clear gains for explorer users despite the higher computational cost.

For inference and production use, we run an offline job that generates ground-truth geo-temporal features and embeddings for the following day. Although these are created one day in advance and may not fully reflect the latest context, they leverage the most recent LLM. As the results indicate, even without up-to-date embeddings at inference time, this setup still achieves the second-best model(M+GT$_{train}$ in Table~\ref{tab:model_comparison production dataset random users}) in terms of relative improvement four our production dataset.



\begin{table*}[t!]
    \centering
    \resizebox{0.68\textwidth}{!}{  
    \begin{tabular}{lccc ccc ccc ccc ccc}
    \toprule
    \textbf{Model Variation} &
    \multicolumn{2}{c}{\textbf{NDCG}} &
    \multicolumn{3}{c}{\textbf{HR}} &
    \multicolumn{3}{c}{\textbf{Coverage}} \\
    \cmidrule(lr){2-3}
    \cmidrule(lr){4-6}
    \cmidrule(lr){7-9}
    \cmidrule(lr){10-12}
    \cmidrule(lr){13-15}
    & \textbf{@5} & \textbf{@10}
    & \textbf{@1} & \textbf{@5} & \textbf{@10}
    & \textbf{@1} & \textbf{@5} & \textbf{@10} \\
    \midrule
    \addlinespace
M & 18.02 & 14.52 & \underline{29.31} & 10.46 & 2.83 & \underline{16.12} & \textbf{72.15} & \underline{109.02} \\
Id+M & -40.91 & -37.21 & -52.49 & -33.44 & -25.31 & -50.62 & -43.35 & -39.16 \\
Id+M+GT$_{train}$ & -18.29 & -16.58 & -24.61 & -14.28 & -10.61 & -37.31 & -35.21 & -33.22 \\
Id+M+GT & -2.11 & -1.54 & -2.50 & -1.90 & -0.60 & 1.40 & 2.69 & 1.65 \\
M+GT$_{train}$ & \underline{19.31} & \underline{15.81} & 28.48 & \underline{13.00} & \underline{5.26} & \textbf{19.33} & \underline{67.96} & \textbf{112.54} \\
M+GT & \textbf{21.44} & \textbf{17.68} & \textbf{32.77} & \textbf{13.90} & \textbf{5.59} & 14.72 & 60.76 & 105.39 \\
Loss-[BCE] & 0.64 & 0.70 & 2.00 & -0.15 & 0.09 & -3.56 & -4.35 & -5.94 \\
Loss-[Cos] & -1.24 & -1.34 & -1.01 & -1.13 & -1.39 & -15.78 & -20.97 & -21.67 \\
Loss-[Pairwise$_{rand}$] & 1.06 & 0.94 & 2.80 & 0.40 & 0.22 & -16.65 & -20.02 & -21.34 \\
Loss-[Pairwise$_{sem}$] & 1.54 & 1.43 & 2.44 & 0.94 & 0.84 & -3.25 & -7.52 & -8.14 \\
\bottomrule
    \end{tabular}
    }
    \caption{Model performance for general users, production dataset. Percentage improvement (\%) over Baseline(SASRec)}
    \label{tab:model_comparison production dataset random users}
    \end{table*}

\begin{table*}[t!]
    \centering
    \resizebox{0.91\textwidth}{!}{%
    \begin{tabular}{lcccccccccc}
    \toprule
    \textbf{Model Variation} & \multicolumn{5}{c}{\textbf{MovieLens Dataset}} & \multicolumn{5}{c}{\textbf{Production Dataset}} \\
    \cmidrule(lr){2-6}\cmidrule(lr){7-11}
    & \textbf{NDCG@5} & \textbf{NDCG@10} & \textbf{HR@1} & \textbf{HR@5} & \textbf{HR@10} 
    & \textbf{NDCG@5} & \textbf{NDCG@10} & \textbf{HR@1} & \textbf{HR@5} & \textbf{HR@10} \\
    \midrule

M & 24.30 & 11.27 & 1250.00 & 9.75 & 0.75 & -72.39 & -75.00 & -73.98 & -72.13 & -76.43 \\
Id+M & 12.91 & 7.10 & 375.00 & 8.40 & 3.32 & -15.95 & -14.45 & -4.07 & -18.64 & -15.56 \\
Id+M+GT$_{train}$ & \textbf{{47.09}} & \textbf{{29.94}} & \textbf{{2337.50}} & \textbf{{21.60}} & \textbf{{12.64}} & 6.44 & 7.80 & 11.38 & 5.27 & \underline{{7.89}} \\
Id+M+GT & \underline{{43.29}} & \underline{{24.07}} & \textbf{{2337.50}} & \underline{{17.28}} & 6.13 & \textbf{{9.51}} & \textbf{{9.63}} & \textbf{{21.95}} & \underline{{6.40}} & \textbf{{8.01}} \\
M+GT$_{train}$ & 23.04 & 12.04 & 1375.00 & 7.78 & 1.63 & -61.35 & -61.47 & -73.98 & -58.19 & -59.84 \\
M+GT & 31.65 & 17.13 & \underline{{1575.00}} & 13.70 & 4.13 & -65.34 & -65.37 & -82.11 & -61.77 & -63.04 \\
Loss-[BCE] & 10.13 & 5.25 & 50.00 & 8.40 & 3.19 & -2.15 & -2.06 & 4.88 & -2.82 & -2.63 \\
Loss-[Cos] & 9.37 & 7.87 & 0.00 & 8.52 & \underline{{7.26}} & -12.27 & -9.17 & -19.51 & -9.98 & -5.61 \\
Loss-[Pairwise$_{rand}$] & 5.82 & 4.01 & 25.00 & 4.69 & 3.13 & -3.37 & -2.75 & -3.25 & -3.01 & -2.52 \\
Loss-[Pairwise$_{sem}$] & -2.78 & -6.79 & 50.00 & -5.80 & -9.64 & \underline{{9.20}} & \underline{{8.03}} & \underline{{17.07}} & \textbf{{7.53}} & 6.29 \\
\bottomrule
    \end{tabular}
    }
    \caption{Model performance comparison for explorer users. Percentage improvement (\%) over Baseline(SASRec). Bold = best, underline = second best per metric within each dataset. Absolute performance for MovieLens is provided in the Appendix.}
    \label{tab:model_comparison_two_datasets explorer users}
    \end{table*}

\subsection{Geo-Temporal Context Adaptation}
Geo-temporal embeddings are designed to capture seasonality, event-driven viewing patterns, and during training they consistently improve performance by providing additional contextual signals. The production data set timestamps reflect the time that the user watches the program. However, the MovieLens dataset contains timestamps reflecting rating events rather than actual watch times, resulting in heterogeneous geo-temporal distributions. This also undermines semantic negative sampling, since candidate negatives may not match the true temporal context of the users. Consequently,  objectives based on binary cross-entropy (BCE) or pairwise losses, that show promising signal in the production dataset results, suffer from misaligned supervision in Movielens results. Whereas cosine similarity loss, relying only on positive pairs, proves more robust and delivers the largest improvements. Nevertheless, geo-temporal patterns learned in training of Movielens dataset may not align with the test context, particularly when certain periods are sparsely represented, introducing noise and bias into ranking. Therefore, we find that omitting geotemporal embeddings in inference, resulting from the Id+M+GT$_{train}$ setting in Table~\ref{tab:model_comparison_two_datasets explorer users}, achieves higher performance than using them throughout training and inference (Id+M+GT) for the Movielens dataset that is in contrast with what is observed for the production dataset. This demonstrates that geo-temporal features are valuable as auxiliary training signals for our production dataset but may inject noise when applied directly at test time for the ML dataset. Overall, our approach offers flexibility: geotemporal embeddings can be selectively applied according to the temporal characteristics of the target data, enabling adaptation to both time-aligned and temporally heterogeneous scenarios.





\vspace{-5pt}
    
\subsection{Qualitative Results}

To better understand model behavior, we examined sample predictions where geo-temporal context improved relevance.

\textbf{Example 1 — \textit{On Golden Pond}}: A user in New Hampshire interacted just after July 4. The summer, outdoor context led the model to recommend a family drama set in a New England lake house.

\textbf{Example 2 — \textit{Sleepless in Seattle}}:  
A user in Seattle interacted just after New Year’s. The holiday-themed context recommended a romantic film set in Seattle.

Although anecdotal, these examples indicate that geo-temporal context can support more nuanced, thematically appropriate recommendations beyond what collaborative signals may offer.

\vspace{-5pt}

\subsection{Conclusion and Future Work}

We introduced a novel approach for enriching recommender systems with LLM-generated geo-temporal context derived from simple signals like timestamp and location. This method produces semantically meaningful features such as holidays, seasonal events, and regional trends that enhance user understanding without relying on long historical data or static knowledge bases. In addition to proposing a geo-temporal context extraction pipeline, we developed a new model integration strategy that incorporates these signals into a sequential recommendation framework. Our experiments on the MovieLens 1M and production dataset show that this approach improves ranking quality and increases recommendation diversity, particularly in contextually distinct scenarios. We also release our enriched dataset to support reproducibility and encourage further exploration of geo-temporal context-aware recommendation. As future work, we plan to evaluate additional datasets and A/B test the propose approach in a production environment with actual user settings.


\bibliography{aaai2026}

\newpage
\appendix
\section{Appendix A. Absolute Performance of MovieLens Dataset}
We have shown explorer results in Table \ref{tab:model_comparison_two_datasets explorer users} using percentage improvment over baseline. Table \ref{tab:model_comparison_ml_explorer} the actual values for each metric are displayed. 

\begin{table}[htbp]
    \centering
    \resizebox{0.7\textwidth}{!}{%
      \begin{tabular}{lccccc}
        \toprule
        \textbf{Model Variation} & \multicolumn{5}{c}{\textbf{MovieLens Dataset}} \\
        \cmidrule(lr){2-6}
        & \textbf{NDCG@5} & \textbf{NDCG@10} & \textbf{HR@1} & \textbf{HR@5} & \textbf{HR@10} \\
        \midrule
        Baseline(SASRec) & 0.0395 & 0.0648 & 0.0008 & 0.0810 & 0.1598 \\
        M & 0.0491 & 0.0721 & 0.0108 & 0.0889 & 0.1610 \\
        Id+M & 0.0446 & 0.0694 & 0.0038 & 0.0878 & 0.1651 \\
        Id+M+GT$_{train}$ & \textbf{0.0581} & \textbf{0.0842} & \textbf{0.0195} & \textbf{0.0985} & \textbf{0.1800} \\
        Id+M+GT & \underline{0.0566} & \underline{0.0804} & \textbf{0.0195} & \underline{0.0950} & 0.1696 \\
        M+GT$_{train}$ & 0.0486 & 0.0726 & 0.0118 & 0.0873 & 0.1624 \\
        M+GT & 0.0520 & 0.0759 & \underline{0.0134} & 0.0921 & 0.1664 \\
        Loss-[BCE] & 0.0435 & 0.0682 & 0.0012 & 0.0878 & 0.1649 \\
        Loss-[Cos] & 0.0432 & 0.0699 & 0.0008 & 0.0879 & \underline{0.1714} \\
        Loss-[Pairwise$_{rand}$] & 0.0418 & 0.0674 & 0.0010 & 0.0848 & 0.1648 \\
        Loss-[Pairwise$_{sem}$] & 0.0384 & 0.0604 & 0.0012 & 0.0763 & 0.1444 \\
        \bottomrule
      \end{tabular}
    }
    \caption{Model performance comparison for explorer users. Bold = best, underline = second best per metric within each dataset.}
    \label{tab:model_comparison_ml_explorer}
\end{table}

\end{document}